\documentclass[a4paper,fleqn,usenatbib]{mnras}

\usepackage{newtxtext,newtxmath}

\usepackage[T1]{fontenc}
\usepackage{ae,aecompl}

\usepackage{graphicx}	
\usepackage{amsmath}	
\usepackage{amssymb}	
\usepackage{multirow}

\usepackage{fancyvrb}
\usepackage[dvipsnames]{xcolor}

\usepackage[utf8]{inputenc}
\usepackage[export]{adjustbox}
\usepackage{wrapfig}

\graphicspath{ {images/} }

\newcommand{\lya}{\mbox{$\rmn{Ly}\alpha$}}

\newcommand{\fesc}{\mbox{$f_{\rm esc}^{\rm Ly \alpha}$}}
\newcommand{\lyacode}{\texttt{FLaREON}} 
\newcommand{\lyart}{\texttt{LyaRT}} 

\title [\lyacode]{\lyacode: a fast computation of \lya\ escape fractions and line profiles}

\author[S. Gurung-L\'opez. et al.]{
Siddhartha Gurung-L\'opez,$^{1}$\thanks{E-mail: sidgurung@cefca.es}
\'Alvaro A. Orsi$^{1}$,
Silvia Bonoli$^{1,2}$
\\
$^{1}$ Centro de Estudios de F\'isica del Cosmos de Arag\'on, Plaza San Juan 1, piso 2, Teruel, 44001, Spain. \\
$^{2}$ DIPC, Manuel Lardizabal Ibilbidea, 4, 20018 San Sebastian, Spain.
\\
}

\date{Accepted XXX. Received YYY; in original form ZZZ}

\pubyear{2018}

\begin{document}
\label{firstpage}
\pagerange{\pageref{firstpage}--\pageref{lastpage}}
\maketitle

\begin{abstract}
We present \lyacode\ (Fast Lyman-Alpha Radiative Escape from Outflowing Neutral gas), a public \texttt{Python} package that delivers fast and accurate \lya\ escape fractions and line profiles over a wide range of outflow geometries and properties. The code incorporates different algorithms, such as interpolation and machine learning to predict \lya\ line properties from a pre-computed grid of outflow configurations based on the outputs of a Monte Carlo radiative transfer code. Here we describe the algorithm, discuss its performance and illustrate some of its many applications. Most notably, \lyacode\ can be used to infer the physical properties of the outflowing medium from an observed \lya\ line profile, including the escape fraction, or it can be run over millions of objects in a galaxy formation model to simulate the escape of \lya\ photons in a cosmological volume.
\end{abstract}




\section{Introduction}

\indent Since the first evidence of star forming galaxies emitting \lya\ photons \citep{steidel96, hu98}, more than two decades ago, observational campaigns targeting these sources have developed to become a standard technique to identify high redshift galaxies \citep[e.g.][]{rhoads00, malhotra02, Konno2016, Sobral2017, Ouchi2018a}. However, we are still far from a comprehensive understanding of this galaxy population due to the complex radiative transfer (RT) processes that \lya\ photons experience \citep[see][for a review]{dijkstra17}. \\

\indent The \lya\ RT in astrophysical media can be addressed analytically for static, simplified geometries \citep[e.g.][]{harrington73, neufeld90}. However, the limited validity of such approach encouraged the development of numerical Monte Carlo radiative transfer (MCRT) codes. In this approach, \lya\ photons are tracked individually as they interact in arbitrarily complex 3D gas geometries. As a result, information about the fraction of photons that manage to escape, \fesc, and their resulting line profile is computed \citep{ahn00, zheng02, ahn03, verhamme06, orsi12, verhamme_2012, laursen09a, Gronke_2016}. 

\indent One drawback of the MCRT technique is given by the time it takes to simulate an appropriate number of photons to
obtain statistically significant results. The mean number of scattering events scales roughly proportionally with the \lya\ rest-frame optical depth of the medium \citep{harrington73}. Hence, the computational time can vary by several orders of magnitude depending on the physical configuration probed. Typically, approximately $10^4-10^5$ photons are needed to retrieve the shape of the \lya\ line profile with reasonable resolution. Such exercise becomes quickly prohibitively expensive when running MCRT codes for multiple configurations. One scenario where this requirement is needed is in \lya\ line profile fitting (e.g. Mejias et al., in prep). 
Another example is to incorporate the \lya\ properties of objects in a galaxy formation model run over a cosmological box \citep[e.g.][]{orsi12, garel12, Gurung_2018}

\indent Here we address the problem described above by presenting \lyacode, a publicly available \texttt{Python} package able to quickly predict multiple Lyman-$\alpha$ line profiles and escape fractions with high accuracy. The basis of the results is a grid of configurations computed previously using the MCRT code \lyart\ \citep{orsi12}.The outline of this work is as follows. In \S 2 we present \lyacode. In \S 3 we test its accuracy and in \S 4 we briefly explain how to exploit \lyacode . In \S 5 we illustrate some possible applications. Finally, we present our conclusions in \S 6.

\begin{figure*} 
\includegraphics[width=6.93in]{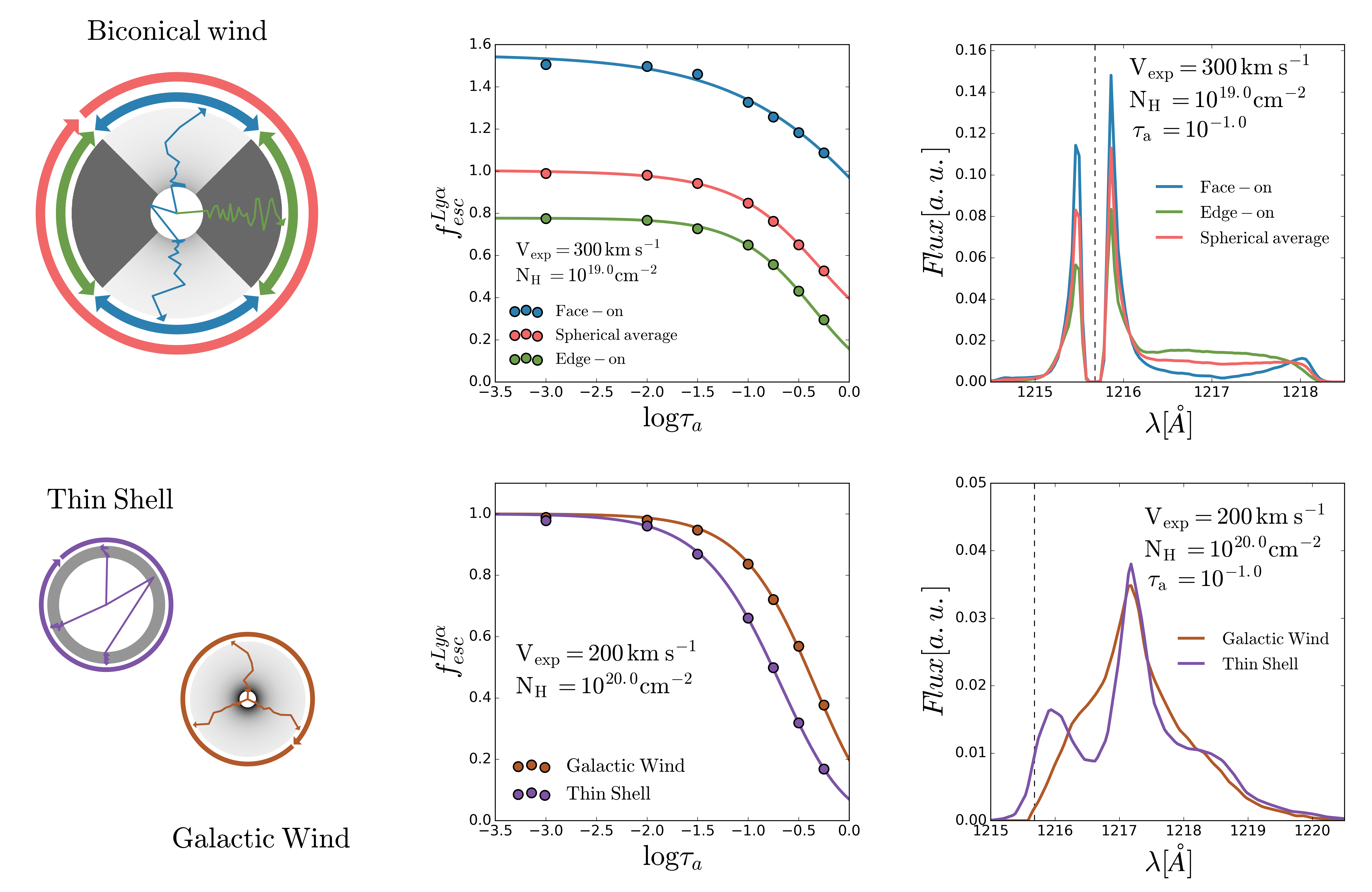} 
\caption{The \lya\ line properties in different geometries. The left panels show a cartoon of the Biconical wind, Thin shell and Galactic wind. The middle panels show the \lya\ escape fraction as a function of dust  optical depth of absorption $\tau_a$ for the three geometries and outflow configuration as shown in the legend. The right panels show
examples of \lya\ line profiles for the configurations described in the legend.}
\label{fig:bicone}
\end{figure*}

\section{Code description}

\indent \lyacode\ \footnote{ \texttt{ https://github.com/sidgurun/FLaREON}} makes use of a grid of configurations run with the Monte Carlo radiative transfer code \lyart \footnote{ \texttt{ https://github.com/aaorsi/LyaRT} } \citep{orsi12}.  This grid covers three different outflow geometries and a wide range of gas properties. The goal of the code is to deliver a user-friendly public \texttt{Python} package able to predict thousands of Lyman-$\alpha$ line profiles and \fesc\ in seconds with minimal user input.

\subsection{Outflow Geometries}\label{ssec:geometries} 

\indent There are three different outflow geometries implemented in \lyacode: {\it Thin shell, Wind,} and {\it Biconical wind}. They all feature an empty inner cavity and an isotropic monochromatic source of \lya\ photons is placed in the centre. Other works in the literature include a non-monocromatic intrinsic \lya\ line profile that produce some modifications in the emerging line profile \citep[e.g.][]{verhamme06}. We plan to implement arbitrary intrinsic line profiles in a future update. In all geometries, dust follows the gas density. The dust optical depth is defined as 
\begin{equation}
\label{eq:recipe-ta}
\centering
\tau_{a} =  ( 1 - A_{{\rm Ly}\alpha} ) \frac{E_{\odot}}{Z_{\odot}}{N_H} Z,
\end{equation}
where $E_{\odot} = 1.77 \times 10^{-21} {\rm cm^{-2}}$ \citep{Silva_1998,orsi12} is the ratio $\tau_a/N_H$ for solar metallicity, $A_{Ly\alpha} = 0.39 $ is the albedo at the Ly$\alpha$ wavelength, $Z_{\odot} = 0.02$  \citep{granato00}, $Z$ is the gas metallicity and $N_{\rm H}$ is the neutral hydrogen column density. We assume the temperature of the gas is constant at $T=10000~{\rm K}$.

\indent The Thin shell and Wind geometries are described in detail in \citet{orsi12} and \citet{Gurung_2018}. In the following we briefly 
describe them. The Biconical wind described below is slightly different to that presented in \citet{Gurung_2018}.

\begin{enumerate}

\item Thin shell: an isothermal uniform neutral hydrogen distributed in a thin layer with a radial expansion velocity ${\rm V_{exp}} $. This geometry has been widely used in the literature to study the escape of \lya\ photons \citep[e.g.][]{zheng02, ahn04, verhamme06, orsi12}.

\item Wind: a spherical isothermal distribution of neutral hydrogen with radial expansion velocity $ V_{\rm exp}$ as implemented in \cite{orsi12}. This geometry exhibits an empty spherical cavity with radius $R_{\rm Wind}$ (analogously to $R_{\rm inner}$ in the Thin shell) and a radially decreasing number density profile.

\item Biconical wind: this is a combination of an outflow with expanding wind geometry and a static isothermal uniform medium. The expanding wind outflow with $ V_{\rm exp,in}>0$ and $ N_{\rm H,in}$ is confined in $\theta<\theta_{\rm cone}$ and $\theta>\pi  - \theta_{\rm cone}$, where $\theta_{\rm cone}$ is measured from the polar axis. We define $\theta_{\rm cone}=\pi/4$. For $\theta_{\rm cone}<\theta<\pi - \theta_{\rm cone}$, the medium is static ($ V_{\rm exp,out}=0$) with column density ${N_{\rm H,out}} = f \times {N_{\rm H,in}}$. Here we arbitrarily set $f=10^3$. This value of $f$ assures that the slab is optically thick in comparison to the bicone, thus most of the photones escape through the bicone. The column density of this geometry $N_{\rm H}$ is 

\begin{equation}\label{eq:bicone-column-density} 
\centering
N_{\rm H} =  (1-\cos \theta _ {\rm cone} ) {N_{\rm H,in}} + \cos \theta _ {\rm cone} {N_{\rm H,out} }
\end{equation} 
where ${N_{\rm H,in}}$ corresponds to the column density of the Wind geometry \citep[see][]{Gurung_2018}. 

\end{enumerate}

\begin{figure} 
\includegraphics[width=3.5in]{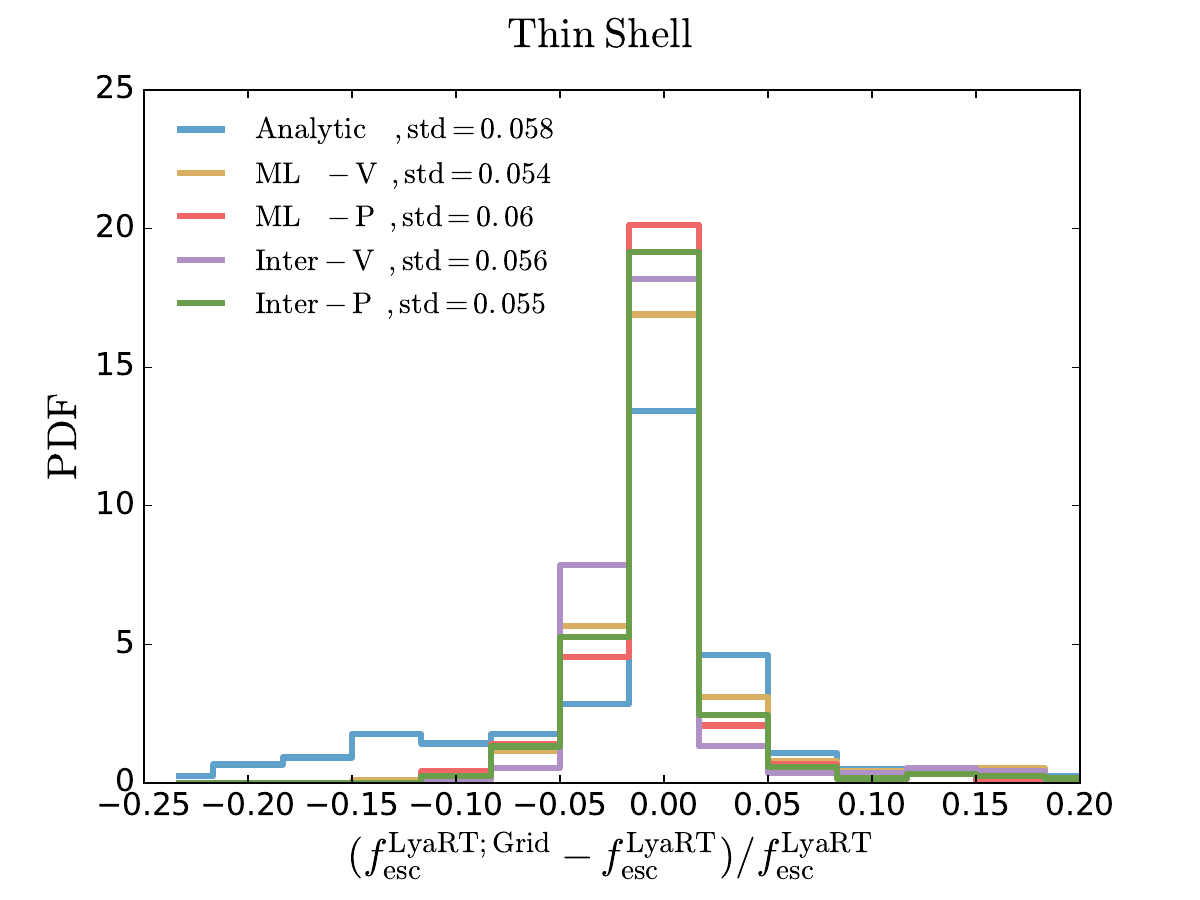} 
\caption{ Distribution of the relative difference between the output of \lyart\ and the predictions of \lyacode\ for 300 random outflow configurations using the Thin shell geometry. The result with the analytical expressions is shown in blue. The result with the machine learning algorithm (labeled as 'ML') using the raw outputs of \lyart\ (labeled as 'V') and using the best fitting parameters of equation \ref{eq:fesc-func} (labeled as 'P') are plotted in red and yellow respectively. Additionally, the algorithms using linear interpolation (labeled as 'Inter') and the raw output of \lyart\ ('V') and using the fitting  parameters ('P') are shown in purple and green respectively. The legend displays the standard deviation of each \lyacode\ algorithm.}
\label{fig:accuracy}
\end{figure}

\indent Fig. \ref{fig:bicone} illustrates the main features of the three geometries described. For the Biconical wind, photons scattering through the dense and static torus are less likely to escape compared to those traveling through the thin outflow. The dipolar nature of the scattering of \lya\ photons through a neutral hydrogen medium makes them likely to back-scatter within the inner cavity until they escape through the thin medium. For $\theta_{\rm cone} = \pi/4$, $\sim 70\%$ of the photons are emitted towards the thick torus and only a few ($\sim 30\%$) towards the bicone.

When treating the biconical geometry, it is important to note that $f_{\rm esc}^{\rm Ly\alpha}$ can be define for an arbitrary angular aperture $\Delta \gamma$ as

\begin{equation}
\centering
f_{\rm esc}^{\rm Ly\alpha} (\Delta\gamma) = {{N_{\rm escaped}(\Delta \gamma)}\over{N_{\rm emitted}(\Delta \gamma)}} 
\end{equation} 

where $N_{\rm emitted}(\Delta \gamma)$ is the number of photons initially sent towards the angle aperture $\Delta \gamma$ and ${N_{\rm escaped}(\Delta \gamma)}$ is the number of photons that manage to escape through that set of directions.  Note that for the non-biconical Galactic wind and the Thin shell this  expression is valid and  $\Delta \gamma$ is every line of sight ($4\pi$). 

The bicone geometry  is not spherically symmetric, thus its $f_{\rm esc}$ depends on the line of sight.  In the case of the face-on bicone, the escape fraction presented in our work corresponds to the $f_{\rm esc}^{\rm Ly\alpha} (\Delta \gamma _b)$ where $\Delta \gamma _b$ is the angular aperture  $\theta<\pi/4 \cup \theta>3\pi/4$ and $\phi \in [0,2\pi)$. In other words, we define the face-on biconical escape fraction as the number of photons that escape through the bicone divided by the number of photons emitted towards the bicone. In a similar way, we define the edge-on escape fraction  as $f_{\rm esc}^{\rm Ly\alpha} (\Delta \gamma _s)$ where $\Delta \gamma _s$ is the angular aperture  $\theta \in [\pi/4,3\pi/4] $ and $\phi \in [0,2\pi)$. 

\indent The inclination of the Biconical wind with respect to the observer leads to different escape fractions. This is shown in the middle panel of Fig.~\ref{fig:bicone} by computing the escape fraction \fesc\ of \lya\ photons though the bicone (face-on) or through the torus (edge-on). The escape fraction is defined as the ratio between the number of photons that escape through a given direction over the number of photons emitted towards that direction. The three cases shown in Fig.~\ref{fig:bicone} (edge-on, face-on and the spherically average) show significant differences. If the geometry is observed face-on, \fesc\ reaches values greater than 1, while if it is edge-on $\fesc < 1$ even if there is no dust. This is caused by the large optical depth of the torus that beams the \lya\ photons towards the bicone. In the spherically averaged \fesc\  reaches unity for dust-free configurations. 

\indent The right panel of Fig. \ref{fig:bicone} shows the resulting line profiles for the bicone at the three different orientations. The differences in the line profile between the three cases reflect the different scattering histories of photons escaping through the outflow or the thick torus.

\indent Fig.~\ref{fig:bicone} also shows the corresponding \fesc\ and line profiles for the Thin shell and Wind, respectively. Here, there is no difference in the escape of \lya\ photons due to the orientation of the outflows. 

\begin{figure*} 
\includegraphics[width=6.93in]{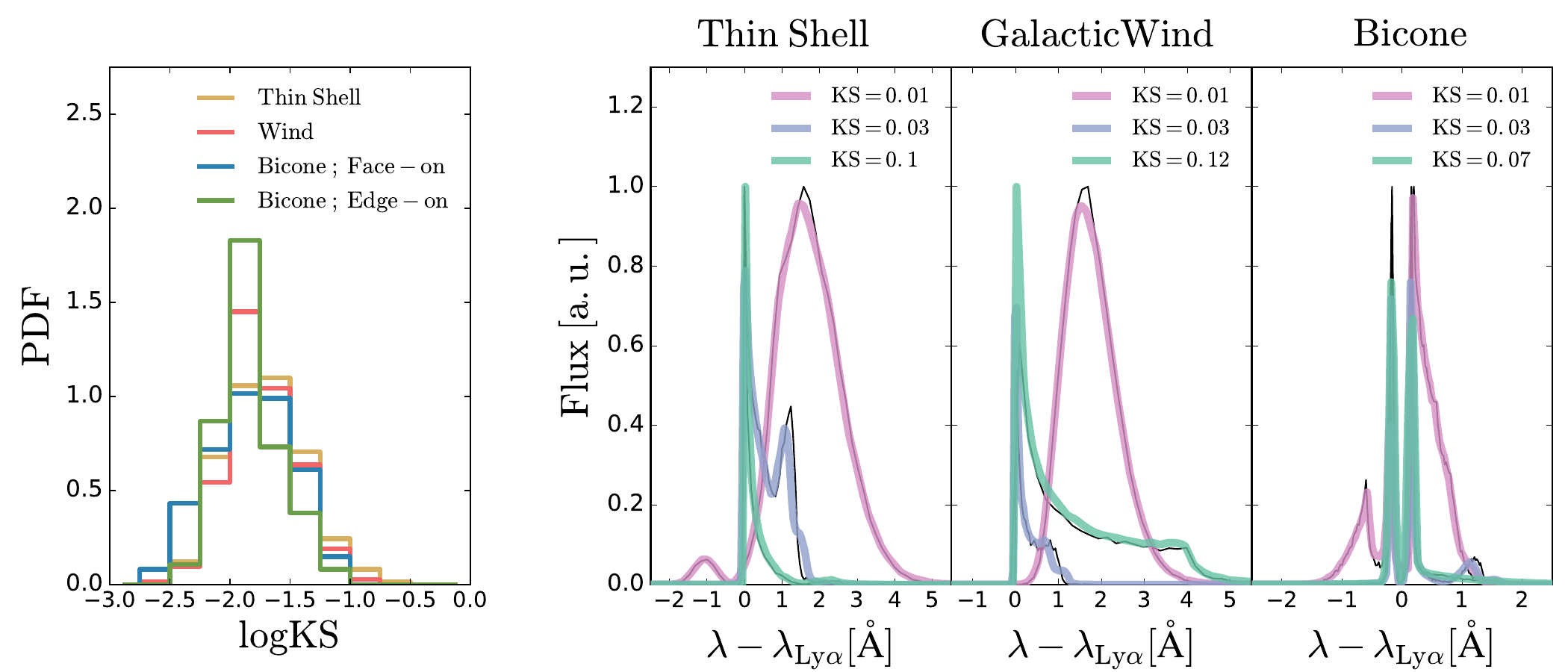} 
\caption{ \textbf{Left} : Normalized distribution of the KS statistic of the \lya\ line profile computed by LyaRT and predicted by \lyacode\ in the same 300 random  \{$\rm V_{exp}$, $\rm N_H$, $\tau_a$\} combinations for the Thin shell (yellow), the Wind (red), and the bicone face-on (blue) and edge-on (green). \textbf{Right} : comparison between the \lyart\ output (solid black lines) and \lyacode\ predictions (colored solid lines) in some of the 300 random configurations. We show Thin shell, Wind and Bicone configurations from left to right. Additionally, in the legend indicates the KS between the \lyart\ and \lyacode\ line profiles. }
\label{fig:KS}
\end{figure*}
   
\subsection{Monte Carlo configuration grids}

\indent \lyacode\ is based in the outputs of \texttt{LyaRT} in a grid of configurations spanning a wide range of phyisical properties. We build two grids for each outflow geometry; one to infer \fesc\ and another to predict \lya\ line profiles. For the \fesc, the grids are constructed using a number of photons $N_p = 10^4$. Hence, the lowest value of \fesc\ computed is $10^{-4}$. For the line profile grids, we use instead $N_p = 10^5$, since we need more photons to fully recover the shape of the resulting line profiles. In order to speed-up the computational time in the latter case, we implement an acceleration procedure to dismiss scattering events that result in frequency changes below a critical value of $x_{\rm crit} = 3$, where $x$ is the frequency of photons in Doppler units \citep[see, e.g.][]{dijkstra06, laursen07, orsi12}. Those scattering events have no significant impact on the resulting line profile, but skipping them can improve the performance of the Monte Carlo calculation by orders of magnitude. Since the bicone geometry is split into two lines of sight we increased the number of photons to $10^5$ and $10^6$ for the \fesc\ and line profile grids, respectively.

\indent The $V_{\rm exp}$-$ \log N_{\rm H}$ parameter space covered in the grids is defined as follows:

\begin{align}
\centering
{V_{\rm exp} [km \; s^{-1}]}  = \left\{
\begin{array}{c l l }
\left[ 10   \; ,   100 \right]    \; , &    { \Delta {V_{\rm exp}} = 10 \; {\rm km \; s^{-1} }} \\
\left[  100 \; ,   1000 \right]   \; , &    { \Delta {V_{\rm exp}} = 50 \; {\rm km \; s^{-1} }} 
\end{array}
\right.
\end{align} 

where $\Delta {V_{\rm exp}}$ is the step between evaluations. Additionally, the neutral hydrogen column density $N_{\rm H}$ is mapped in bins of $\Delta \log {N_{\rm H}}= 0.25$ and spans the following range:

\begin{align}
\centering
{\log N_{\rm H} {\rm [cm^{-2}]}}  = 
\left[ 17   \; ,   22 \right]    \; \; , \; \; { \Delta { \log N_{\rm H}[{\rm cm^{-2}]}} = 0.25 }
\end{align} 


\indent The values of dust optical depth $\tau_{\rm a}$ where the grid is sampled are different for the \fesc\ and line profile grids. We have checked that a sparse sampling of \fesc\ is sufficient to deliver good results. These values are
$\log \tau_{\rm a} = \left[ -3.0  \; , -2.0  \; , -1.5 \; , -1.0 \; , -0.75 \;, -0.5 \; , -0.25 \; , \; 0.0 \right]$.
\indent The line profile grids, on the other hand, spans a wider range and more frequently in $\tau_{\rm a}$ to track properly the evolution with dust optical depth of the \lya\ line profile. We cover the low dust range with higher density as we found that the evolution is stronger in this range. Finally, the dust optical depth values are

\begin{equation}
\centering
{\log \tau _{\rm a} }  = \left\{
\begin{array}{c l l }
\left[ -3.75  \; ,   -1.500 \right]    \; , &    { \Delta \log \tau _{\rm a} = 0.25  } \\
\left[  -1.50  \; ,   -0.125 \right]   \; , &    { \Delta \log \tau _{\rm a} = 0.125 } \\
\end{array}
\right.
\end{equation} 
where $\Delta \log \tau _{\rm a}$ is the step between evaluations.\\

\indent All together, the total number of configurations sampled for each geometry is 4704 and 12348 for the \fesc\ and line profile grids, respectively.

\subsection{Predicting the Lyman-$\alpha$ properties. }

\indent \lyacode\ allows the user to choose among three different methods to compute the \fesc from the outflow properties:

\begin{figure*} 
\includegraphics[width=6.93in]{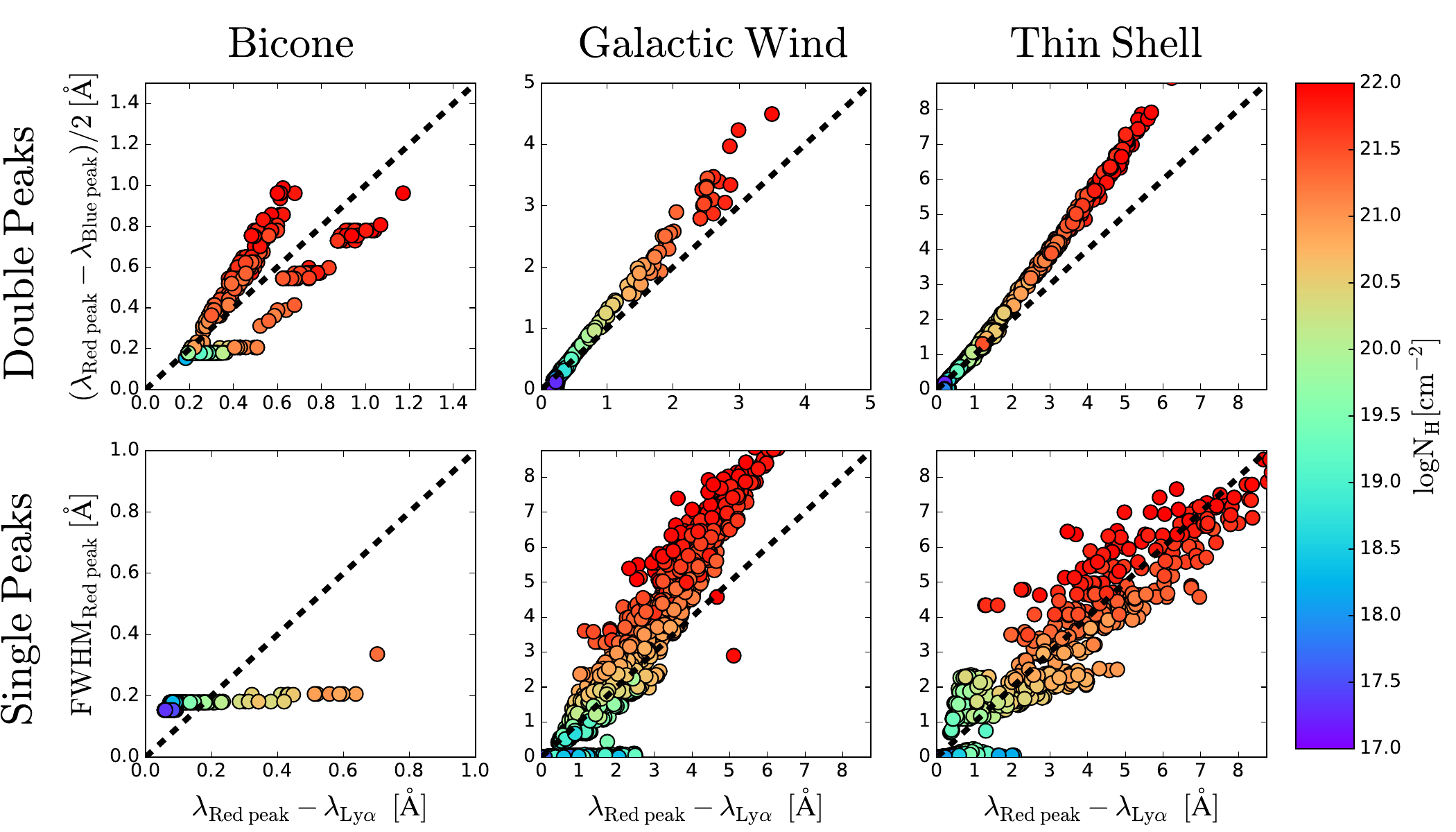} 
\caption{ Correlation between some line profile properties produced by the Bicone, Galactic wind and Thin shell geometries from left to right. \textbf{Top} :   half of the separations between the red and the blue peaks in the line profile as a function of shift of the red peak from \lya\ wavelength. Only configurations producing double peak profiles were taken into account for this panel. \textbf{Bottom} :   full width half maximum of the red peak as a function of the shift of the red peak from \lya\ wavelength . Only configurations producing line profiles without a blue peak were taken into account for this panel. In every panel, the $\rm N_H$ is color coded. The one-to-one relation is plotted in black dashed lines to guide the eye. }
\label{fig:verhamme}
\end{figure*}

\begin{enumerate}
    
\item{ \textbf{ The first method} uses directly the values of \fesc\ given by \lyart\ to build a multi-dimensional linear interpolation grid or to train a machine learning algorithm. For the latter, \lyacode\ incorporates 'extra trees', 'random forest' and 'k-nearest neighbors', using the \texttt{Python} module \texttt{scikit-learn} \citep{scikit-learn}. \\}

\item{ \textbf{ The second method} consists in using a parametric equation that links \fesc\ and one or several gas properties. In particular, we fit the \fesc computed by \lyart\ as a function of the dust optical depth in each node of the $V_{\rm exp}$-$ \log N_{\rm H}$ space to the function inspired by \cite{neufeld90} :

\begin{equation}\label{eq:fesc-func}
\centering
f_{\rm esc}^{\rm Ly\alpha} = k_3 \left[\cosh\sqrt{ k_1 \tau_{\rm a}^{k_2}}\right]^{-1},
\end{equation}    
where $k_1$, $k_2$ and $k_3$ are free parameters. Note that $k_3 = 1 $ in the Thin shell and Wind geometries since if there is no dust $f_{\rm esc}^{\rm Ly\alpha}=1$. However, in the Bicone geometry, as we divide in edge-on and face-on, in general, $k_3 \neq 1 $ as discussed in \S\ref{ssec:geometries}. The best fitting parameters are then used to build the linear interpolated grid or to train the machine learning algorithms in this mode. \\}

\item{ \textbf{The third method} consists on an updated version of the analytic \fesc\ expressions described in \citet{Gurung_2018}. Those \fesc\ parametric equations have been recalibrated for the Thin shell and Wind in our grid, which is denser and wider in $V_{\rm exp}$ and $N_{\rm H}$. Due to the complexity of \fesc\ in the face-on and edge-on configurations, \lyacode\ does not include any analytic expression for the biconical outflow.\\}
\end{enumerate}

\indent To address the problem of predicting \lya\ line profiles, analytic expressions and machine learning algorithms perform poorly compared to a multi-dimensional linear interpolation of the \lyart\ outputs. Hence, unlike the case of the \fesc, the \lya\ line profiles are computed using only a multi-linear interpolation of the grids.

\section{Validation of the code}

\indent Here we compare the performance of different methods to obtain \fesc. We compute 300 random values for $V_{\rm exp}$, $\log N_{\rm H}$ and $ \log \tau_{\rm a}$ using a Latin hypercube algorithm to populate randomly and homogeneously the three dimensional space within the range covered by the grids. Fig. \ref{fig:accuracy}  shows the relative difference between the \lyacode\ predicted \fesc\ and the \lyart\ output in those 300 random configurations using the Thin shell geometry. We find a remarkably good match between \lyart\ and \lyacode . The analytic functional form has worse performance (although $\sim 70\%$ is above $90\%$ accuracy) since these were optimized to a smaller $V_{\rm exp}$-$N_{\rm H}$ region.  Additionally, $\sim 80\%$ of the configurations using directly \fesc\ from \lyart\ to train or to interpolate have an accuracy better than $90\%$. The method that gives the best results is the parametric interpolation, as $\sim 95\%$ and $\sim 50\%$ of the configurations have relative differences below 0.1 and 0.01. Other outflow geometries perform similarly.

\indent To quantify the performance of \lyacode\ predicting \lya\ line profiles we perform a Kolmogorov–Smirnov (KS) test over the  300 random outflow configurations for each geometry.Fig.\ref{fig:KS} shows the resulting KS distributions of the 300 random configurations for all the geometries. We find a very good agreement between the \lya\ line profiles computed by \lyart\ and those predicted by \lyacode. For all geometries the KS distribution peaks around ${\rm KS}=10^{-2}$, implying that the typical maximum difference of the cumulative line profiles is $\approx 1\%$ of the flux. Additionally, about $90\%$ of the random samples exhibit ${\rm KS}<0.05$ for all geometries. 

The right panels in Fig.~\ref{fig:KS} show some example geometries where the outputs of Monte Carlo \lyart\ and \lyacode\ are compared. Overall, the differences between the Monte Carlo code and \lyacode\ are negligible.

\begin{figure*} 
\includegraphics[width=6.93in]{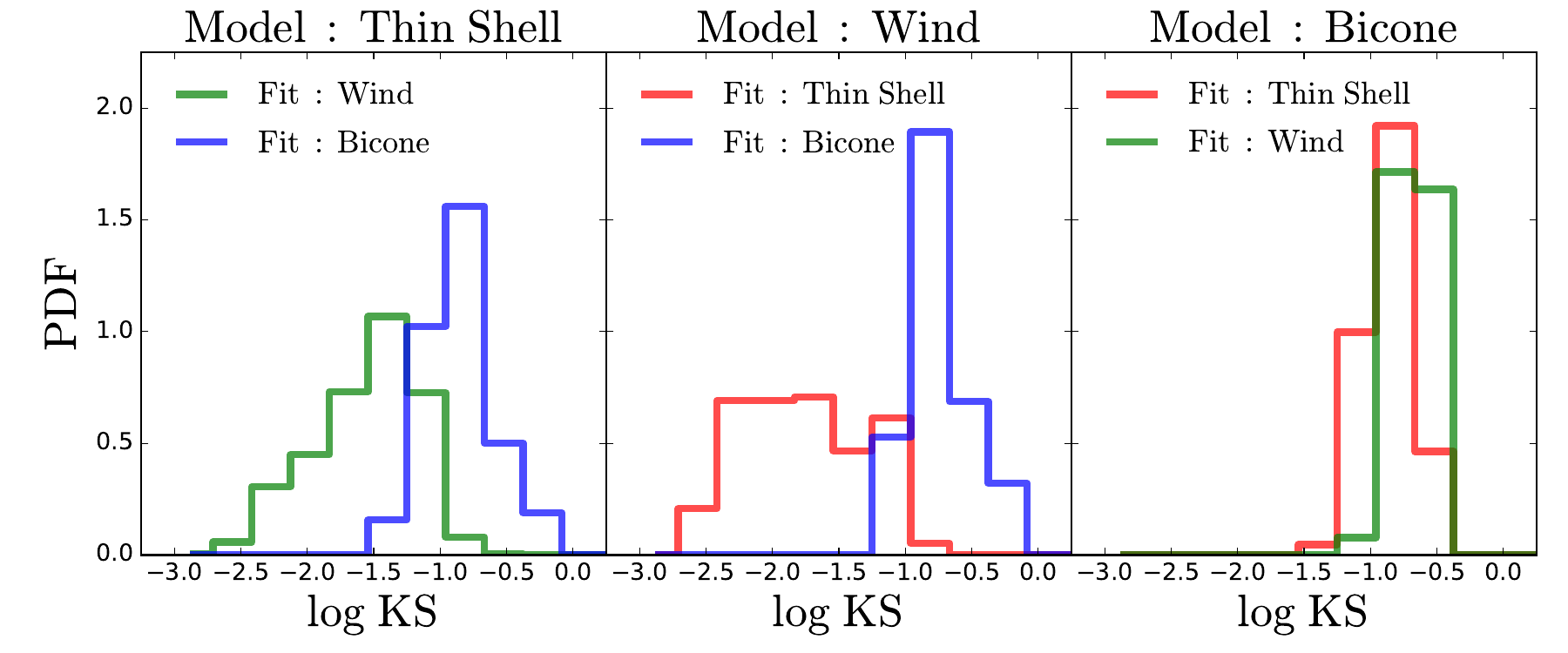}
\caption{ Kolmogorov-Smirnov estimator distribution for the 1000 best fitting line profiles in different geometries. From left to right, the  gas geometries used as {\it models} are: Thin shell, Galactic wind and Biconical wind. The geometries used for the fitting in each panel are the two complementary to the one used for the {\it model}. The Thin shell  is shown in red, the Wind in green while the Bicone is printed in blue.}
\label{fig:performance}
\end{figure*}

\section{Hands on \lyacode}

\indent In this section we ilustrate how to execute \lyacode. After the installation, running \lyacode\ should only take a few \texttt{Python} command lines. 

\indent A simple script to compute the \lya\ escape fraction and line profiles for some given Thin shell configurations is given below:

\begin{Verbatim}[commandchars=\\\{\}]
\textcolor{Mulberry}{import} FLaREON as Lya, numpy as np, pylab as plt
\textcolor{BlueViolet}{# 1) We define the configuration parameters.}
\textcolor{BlueViolet}{# 1.1) Expansion velocity in km/s :}
V_exp = [ \textcolor{BrickRed}{50} , \textcolor{BrickRed}{100} , \textcolor{BrickRed}{200} , \textcolor{BrickRed}{300} ]
\textcolor{BlueViolet}{# 1.2) Logarithm of column densities in cm**-2 :}
log_NH= [ \textcolor{BrickRed}{18} , \textcolor{BrickRed}{19} , \textcolor{BrickRed}{20} , \textcolor{BrickRed}{21} ]  
\textcolor{BlueViolet}{# 1.3) Dust optical depths :}
tau_a  = [ \textcolor{BrickRed}{0.5} , \textcolor{BrickRed}{0.1} , \textcolor{BrickRed}{0.05} , \textcolor{BrickRed}{0.01} ] 
\textcolor{BlueViolet}{# 1.4) Select a geometry :}
Geometry = \textcolor{BrickRed}{'Thin_Shell'}
\textcolor{BlueViolet}{# 2) Compute the escape fractions.}
f_esc = Lya.RT_f_esc( Geometry, V_exp , 
                      log_NH  , tau_a ) 
\textcolor{BlueViolet}{# 3) Compute the line profiles.}
\textcolor{BlueViolet}{# 3.1) Define the wavelength range in meters.}
wavelength = np.linspace( \textcolor{BrickRed}{1213e-10}, \textcolor{BrickRed}{1224e-10}, \textcolor{BrickRed}{1000} )
\textcolor{BlueViolet}{# 3.2) Execute FLaREON}
lines = Lya.RT_Line_Profile( Geometry, wavelength, 
                             V_exp   , log_NH    , 
                             tau_a               )
\textcolor{BlueViolet}{# 4) Display the line profiles.}
\textcolor{BurntOrange}{for} line \textcolor{BurntOrange}{in} lines :
    plt.plot( wavelength , line )
plt.show()

\end{Verbatim}

\indent Other examples can be found in the GitHub repository, including the coupling of \lyacode\ with other popular public codes such as \texttt{emcee}\footnote{\url{http://dfm.io/emcee/current/}} \citep{emcee} to perform \lya\ line fitting. 

\begin{figure*} 
\includegraphics[width=6.93in]{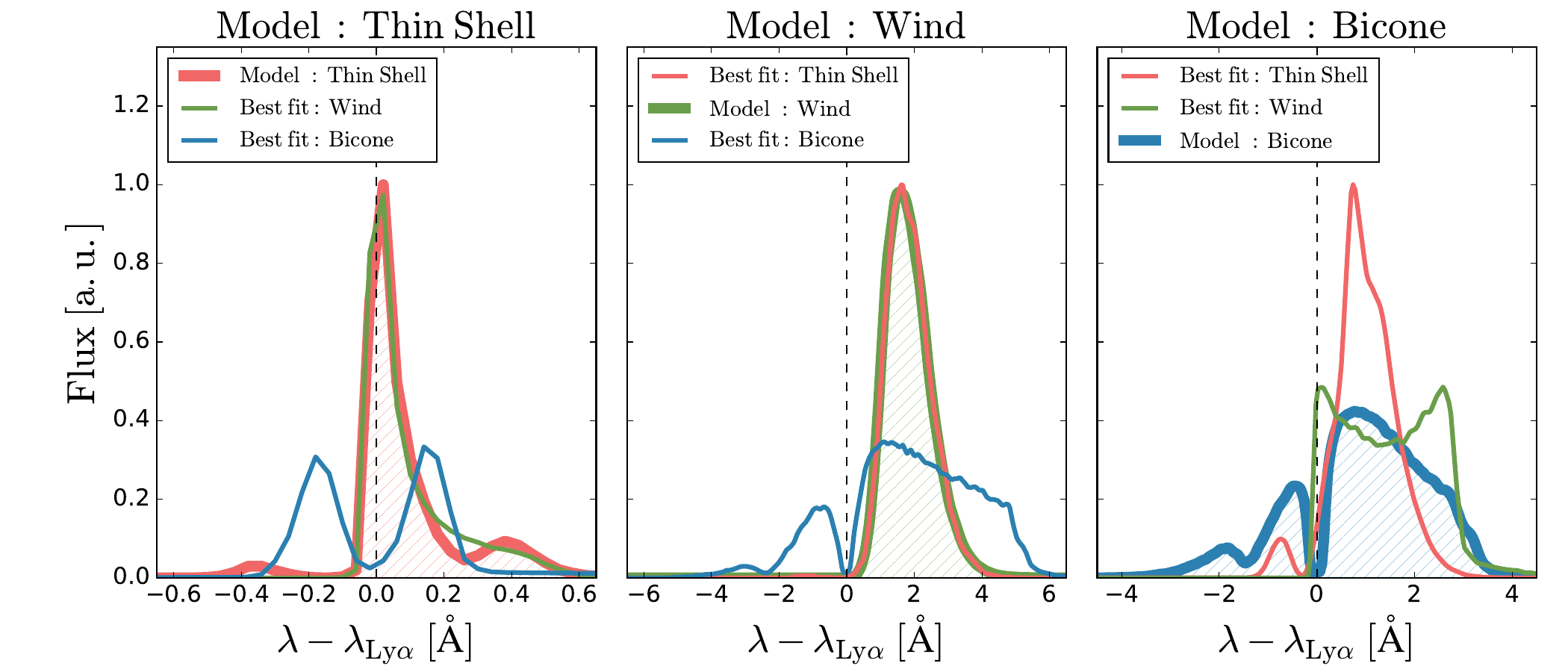} 
\caption{ Comparison of the randomly selected model's \lya\ line profiles and the best fitting outflow configurations using different gas geometries. The Thin shell, Galactic wind and Bicone are shown in red, green and blue respectively. Additionally line profile used as model is highlighted by thicker lines. The model outflow geometry changes from Thin shell, to Galactic wind to Bicone from left to right.}
\label{fig:mcmc}
\end{figure*}

\section{Some applications}

\indent In this section we present a small glimpse of the potential scientific applications of \lyacode. \\

\subsection{Line profile properties}

\indent Recently, \cite{Verhamme_2018} characterized the \lya\ line profiles to infer their displacement redwards of the line centre, allowing to infer the systemic redshift of a source from the \lya\ line only. This is done by measuring the difference in wavelength between different peaks, if there is more than one, or by measuring the FWHM of a single peak.

Fig.~{\ref{fig:verhamme}} shows the relation between different line profile properties for 1000 different configurations spanning the full $V_{\rm exp} - N_{\rm H} - \tau_{\rm a}$ space using the three geometries. In the top panels we show the relation between the red peak shift and half of the distance between the red and blue peak in only configurations exhibiting both peaks. In general, \lyacode\ predicts a tight correlation between these properties and the neutral hydrogen column density. We find that \lyacode\ reproduce the correlation found by \cite{Verhamme_2018} for shift of the read peak smaller than 2\AA\ where trend leaves the one-to-one relation and the slope increases. 

\indent In the bottom panels of  Fig.\ref{fig:verhamme}  we show the relation of the shift and the FWHM of the red peak in configurations where only the red peak was found. Overall, \lyacode\ also predicts a correlation between these properties and the column density, although with a greater scatter. Our results with \lyacode\ are consistent with the findings of \citet{Verhamme_2018} for the Thin shell and Galactic wind geometry.

%
%

\begin{table}
\centering
\caption{Description of randomly selected configurations ({\it models}) and their best fitting configuration in other outflow geometries. }
\label{tab:mcmc}
\begin{tabular}{lccccc}
\multicolumn{1}{c}{Configurations}      & $V_{\rm exp}$        & $\log \; N_{\rm H} $  & $\rm \log \; \tau_a$  & $\rm log \; KS $  & $f_{\rm esc}$   \\
\multicolumn{1}{c}{}                    & $\rm [km/s]$         & $\rm [cm^{-2}]$      & \multicolumn{1}{l}{}  &                   &                  \\ \hline
\multicolumn{1}{l}{}                    & \multicolumn{1}{l}{} & \multicolumn{1}{l}{} & \multicolumn{1}{l}{}  &                   &                  \\
Model    \hspace{0.04cm}   : Thin shell & 55.0          & 17.62          & -0.23           & -       & 0.43 \\
Best fit                   : Wind       & 62.4          & 17.14          & -0.25           & -1.5       & 0.58 \\
Best fit                   : Bicone     & 890.5          & 17.11          & -0.13           & -0.5       & 0.46 \\
\multicolumn{1}{l}{}                    & \multicolumn{1}{l}{} & \multicolumn{1}{l}{} & \multicolumn{1}{l}{}  &                   &                  \\ \hline
\multicolumn{1}{l}{}                    & \multicolumn{1}{l}{} & \multicolumn{1}{l}{} & \multicolumn{1}{l}{}  &                   &                  \\
Model    \hspace{0.04cm}   : Wind       & 123.0          & 20.25          & -1.29           & -       & 0.88 \\
Best fit                   : Thin shell & 93.2          & 20.5          & -0.29           & -2.2       & 0.07 \\
Best fit                   : Bicone     & 616.3          & 22.0          & -1.38           & -0.6       & 0.8 \\
\multicolumn{1}{l}{}                    & \multicolumn{1}{l}{} & \multicolumn{1}{l}{} & \multicolumn{1}{l}{}  &                   &                  \\ \hline
\multicolumn{1}{l}{}                    & \multicolumn{1}{l}{} & \multicolumn{1}{l}{} & \multicolumn{1}{l}{}  &                   &                  \\
Model    \hspace{0.04cm}   : Bicone     & 375.0          & 21.42          & -2.37           & -       & 0.97 \\
Best fit                   : Thin shell & 93.0          & 19.64          & -2.75           & -0.8       & 0.99 \\
Best fit                   : Wind       & 348.2          & 19.5          & -1.75           & -0.7       & 0.98
\end{tabular}
\end{table}

\subsection{Extract outflow information from \lya\ line profiles.}

One of the most attractive application of \lyacode\ consists in inferring outflow properties from measured \lya\ line profiles. Usually, in this kind of analysis only one outflow geometry is implemented \citep[e.g.][]{Orlitov_2018,Gronke_2017}. However, since \lyacode\ includes several gas configurations the analysis can be extended to different outflow geometries. In this section we give a glimpse of the advantages of using several geometries and we will exploit further this idea in an upcoming work (Mejias et. al., in prep).

\indent To study how the inferred outflow properties depend on the gas geometry we start by generating a \lya\ line profile with a given gas geometry and random outflow parameters. We refer to this as {\it model}. Then, we perform an MCMC analysis combining \lyacode\ and \texttt{emcee} \citep{emcee} with the other gas geometries to fit the line profile of the {\it model}.

\indent We repeat this test 1000 times per gas geometry.  For the biconical geometry we use the spherically averaged configuration, but we have checked that the results resemble those obtained with the edge-on and face-on configurations. We show the Kolmogorov-Smirnov estimator distribution of the 3000 tests in Fig.\ref{fig:performance}. In general, we find that the Thin shell geometry is able to produce very similar line profiles to the Galactic wind and vice verse. For both geometries we find that, when fitting the other geometry, the typical KS values fall below 0.1. In contrast we find that the line profiles produced by the biconical wind are very different from those generated by the other two geometries. In fact we find that neither of the other geometries is able to procure good fittings to the biconical geometry and vice verse. Indeed, most of the best fitting exhibit KS$>0.1$. 

Additionally, we also show three detailed cases, one per gas geometry. These few examples illustrate the degeneracy between gas geometry and outflow parameters. In Fig. \ref{fig:mcmc}  we show a comparison between the {\it model}’s \lya\ line profile and the best fits of the other outflow geometries, whereas in Table \ref{tab:mcmc} we summarize the results of the MCMC. On one hand, the morphology of the \lya\ line profiles produced by the Bicone are very different ($\rm KS>0.1$) from the line profiles generated by the Wind and Thin shell. Hence, there is not confusion between these gas geometries. On the other hand, the Thin shell and Wind line profiles resemble and fit each other \lya\ line profile with very good agreement ($\rm KS<0.1$). However, since the gas morphology impacts the resulting line profile, the inferred properties from the fit, generally, do not match the {\it model}’s characteristic. 

\indent Additionally, the \fesc\ computed from the inferred outflow properties differs from the \fesc\ of the {\it model}. This increases the difficulty of calculating the intrinsic \lya\ flux emitted before the radiative transfer processes by using only \lya\ emission. 

In a similar fashion, \lyacode\ can also be used to extract information from inflows. The \lya\ line profile emerging from an inflow with a given set of properties (gas topology, $N_{\rm H}$, $\tau_{\rm a}$ and $V_{\rm exp}<0$) can be directly computed from the outputs of \lyacode. Indeed, it is as simple as i) computing the emerging line profile of an outflow with the same gas geometry and set of properties  \{$N_{\rm H}$, $\tau_{\rm}$, $ |V_{{\rm exp}}|$\} and ii) mirroring the line profile with respect to the \lya\ wavelength, i.e., making the mapping 
\begin{equation}
    \centering
     \lambda - \lambda_{\rm Ly\alpha} \rightarrow -(\lambda - \lambda_{\rm Ly\alpha}).
    \label{eq:mapping}
\end{equation}

\begin{figure} 
\includegraphics[width=3.3in]{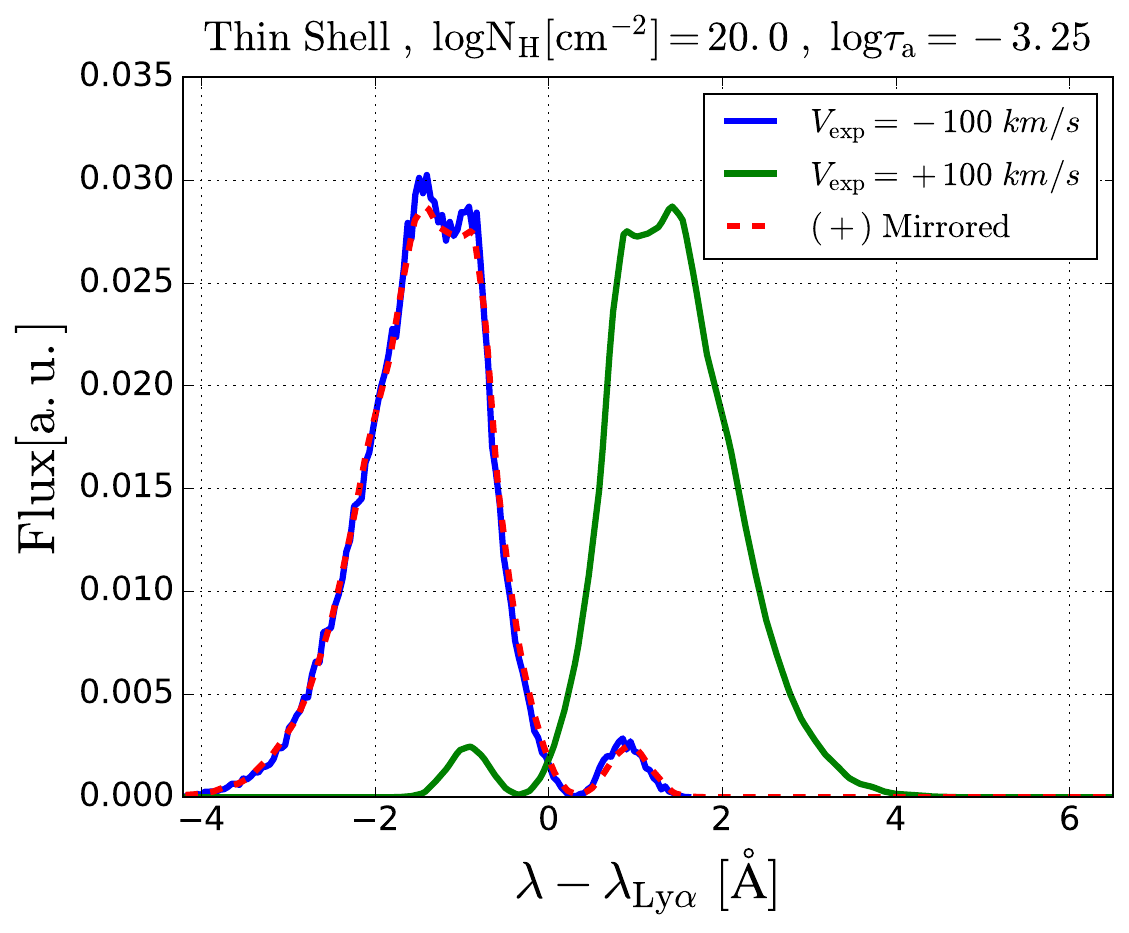} 
\caption{ Comparison between the proper calculation of a \lya\ line profile emerging from an inflow configuration using \lyart\ and the outputs of \lyacode . This example was built with the Thin shell as gas geometry, $\log N_{\rm H}{\rm [cm^{-2}]}=20$, $\rm \log \tau_{\rm a} = -3.25$ and ${ |V_{\rm exp}|=100}\ {\rm km/s}$. In blue we show the full RT computation by \lyart\ using ${V_{\rm exp}=-100}\ {\rm km/s}$. In green we show the \lyacode 's prediction using ${V_{\rm exp}=+100}\ {\rm km/s}$. In dashed red we show \lyacode 's output after the remapping of the wavelength (Eq.\ref{eq:mapping}).}
\label{fig:mirrored}
\end{figure}

In order to validate this procedure we have run \lyart\ in several configurations with $\rm V_{exp}<0$. Then, for the same gas geometry, $\rm \log N_H$, $\rm \log \tau_a$ and $\rm |V_{exp}|$, we generate \lya\ line profiles with \lyacode\ and we apply the methodology explained above to them.  In Fig.\ref{fig:mirrored} we show a comparison  between the proper calculation of the \lya\ line  profile  emerging  from  an  inflow  configuration  using \lyart\ and  the  outputs  of \lyacode\ . After the wavelength remapping, the output of \lyacode\ reproduces very well the full RT calculation. We have checked that this procedure works over the full parameter space exploredby \lyacode .

\section{Conclusions}

\indent In this work we have introduced \lyacode, a user-friendly public \texttt{Python} code based on the radiative transfer Monte Carlo code \lyart\ \citep{orsi12}. This code is able to predict \lya\ line profiles and \lya\ escape fractions for different outflow geometries in a wide range of outflow properties without the need to run a Monte Carlo code.

\indent  \lyacode\ includes three different outflow geometries, an expanding Thin shell \citep[e.g.][]{verhamme06,orsi12, Gurung_2018}, Galactic wind \citep{orsi12, Gurung_2018}, and a  biconical outflow surrounded by a very thick static torus.

\indent In order to predict the \lya\ line profile and \fesc\, \lyacode\ interpolates or makes use of machine learning algorithms over previously-computed grids of configurations. This grids are very dense and cover a wide range of $\rm V_{exp}$, $\rm N_H$ and $\tau_a$. These are composed by 4704 and 12348 different outflow configurations to predict the \fesc\ and line profile, respectively.  

\indent We have analyzed the performance of \lyacode\ of the different geometries and predicting method implemented. \lyacode\ is able to predict thousands of \fesc and line profiles with remarkably high accuracy. The error in \fesc\ is typically below $5\%$. Additionally, the Kolmogorov-Smirnov test delivers values about 0.01 between the fully computed \lyart\ line profiles and \lyacode\ predictions.

\indent In future works we plan to exploit the \lyacode\ capabilities to predict thousands of line profiles and escape fractions to extract outflow physical information such as $V_{\rm exp}$ or $N_{\rm H}$ from observed spectra and to populate large cosmological volumes with \lya\ emitters \cite[see][]{gurung_2019}.

\section*{Acknowledgements}

\indent We thank CEFCA's scientific staff for their useful comments and help, in particular during the debugging phase. The authors acknowledge the support of the Spanish Ministerio de Economia y Competividad project No. AYA2015-66211-C2-P-2. AO and SGL  acknowledge funding from the European Union’s Horizon 2020 research and innovation programme under grant agreement No 734374.




\bibliographystyle{mnras}
\bibliography{ref} 







\bsp	
\label{lastpage}
\end{document}